\documentclass[twocolumn,showpacs,pre,floatfix,letter]{revtex4-1}

\ifx\pdfoutput\undefined
\usepackage{graphicx}
\else
\usepackage{epstopdf}
\usepackage{epsfig}
\usepackage{amssymb,amsmath,stmaryrd,tabularx}
\usepackage{wrapfig}
\usepackage{bm}
\fi

\usepackage[center]{subfigure}

\begin{document}
\title{Path integral calculation for emergence of rapid evolution from
demographic stochasticity}
\author{Hong-Yan Shih and Nigel Goldenfeld}
\affiliation{Department of Physics, Center for the Physics of Living
Cells and Institute for Genomic Biology, University of Illinois at
Urbana-Champaign, Loomis Laboratory of Physics, 1110 West Green Street,
Urbana, Illinois, 61801-3080}


\begin{abstract}
Genetic variation in a population can sometimes arise so fast as to
modify ecosystem dynamics.  Such phenomena have been observed in
natural predator-prey systems, and characterized in the laboratory as
showing unusual phase relationships in population dynamics, including
a $\pi$ phase shift between predator and prey (evolutionary cycles) and
even undetectable prey oscillations compared to those of the predator
(cryptic cycles).  Here we present a generic individual-level
stochastic model of interacting populations that includes a
subpopulation of low nutritional value to the predator. Using a master
equation formalism, and by mapping to a coherent state path integral
solved by a system-size expansion, we show that evolutionary and
cryptic quasicycles can emerge generically from the combination of
intrinsic demographic fluctuations and clonal mutations alone, without
additional biological mechanisms.

\end{abstract}
 \maketitle


Predator-prey ecosystems exhibit noisy population oscillations whose origin
is intuitively quite clear. The predator population number is activated by
the prey, and so increases.  This in turn inhibits the growth of the prey
population, but the decline of the prey leads to a corresponding decline in
the predator number too.  As a result the prey population begins to rise, and
the cycle begins again.  The simplicity of this narrative belies the
difficulty of making a quantitative model of ecosystems. Strong demographic
fluctuations degrade the utility of population-level modeling, rendering it
problematic to assess the appropriate scales for ecological modeling
\cite{levin1992,bascompte1995rethinking,pascual1999individuals,pascual2001,
goldenfeld2011,chave2013} and even influencing community assembly on evolutionary time scales \cite{murasePRE2010}. For example, observations of noisy periodicity in
time series \cite{elton1942}, slowly-decaying correlations
\cite{pinedakrch2007} and spatiotemporal patterns
\cite{bonachela2012patchiness} clearly reflect the stochastic nature of
populations \cite{bonsall2004demographic,deangelis2005individual} and their
spatial organization.  Moreover, even the simplest predator-prey systems
exhibit complex spatial structure.  This can arise through a variety of
pattern formation processes
\cite{meron2012pattern,liebhold2004spatial,malchow2008mathematical,
hillerislambers2001vegetation, blasius1999complex} that include recent
results on deterministic \cite{levin1976hypothesis,kinast2014interplay} and
fluctuation-induced Turing instabilites \cite{butlerPRE2009b,
butlerPRE2011,bonachela2012patchiness}, traveling
waves \cite{sherratt1995ecological,blasius1999complex,mobilia2007} and even
analogies to the processes of phase separation in binary alloys
\cite{liuPNAS2013}. In short, collective and stochastic many-body phenomena
are ubiquitous in biology, and perhaps nowhere more so than in ecology.

The classical literature on predator-prey systems
\cite{berryman1992orgins} assumes that evolution occurs on such long
time scales that it can be neglected, but it is not obvious that this
is always valid \cite{thompson1998}. Recent work using rotifers
(predator) and algae (prey) in a chemostat shows that dramatic changes
in the population structure of the rotifer-algae predator-prey system
can arise from rapid responses to intense selection among induced
genetically distinct strains
\cite{fussmann2000,yoshida2003,yoshida2007,jones2007,becks2010,becks2012,bohannan1997,ellner2013}.
In these studies, so-called sub-populations with different traits
emerge from evolution and lead to new trophic structures, accompanied
by anomalous ecological dynamics.  These anomalies include \lq
evolutionary cycles\rq\, with long oscillation periods in population
dynamics and predator-prey phase shifts near $\pi$ (and definitely
distinct from the canonical value of $\pi/2$), and \lq cryptic
cycles\rq , in which prey populations remain almost constant while the
predator population oscillates. Such phenomena have been modeled with
deterministic differential equations containing empirical descriptions
of functional response with a variety of detailed hypotheses on the
mechanism of species interactions for rapid evolution
\cite{bohannan1997,shertzer2002,yamauchi2005,yoshida2003,
yoshida2007,jones2007,cortez2010,mougi2012,mougi2012b} or non-heritable
phenotypic plasticity  \cite{yamamichi2011}. Such models are not only
very complex, with many adjustable parameters, but also cannot capture
the stochasticity evident in the observations.

The purpose of this Rapid Communication is to propose and analyze a minimal model for rapid
evolution that includes the effects of demographic stochasticity. Using
tools from statistical mechanics, demographic stochasticity has been
successfully captured using individual-level models (ILM) in a variety
of situations that range from simple well-mixed predator-prey
interactions \cite{mckanePRL2005,morita2005,black2012stochastic} to
spatially-extended systems that can exhibit quasi-Turing patterns
\cite{mobilia2007,lugo2008quasicycles,butlerPRE2009,butlerPRE2009b,butlerPRE2011,tauber2012}.
Here we propose an ILM for rapid evolution that we solve
analytically by mapping the model into a coherent-state path integral
representation
\cite{doi1976,grassberger1980fock,mikhailov1981path,goldenfeld1984kinetics,peliti1985path} (for a review and
history, see Ref. (\cite{mattis1998uses})) followed by a volume
expansion \cite{kampen1961power} to derive the effective Langevin
equation for demographic fluctuations. Accompanied by Gillespie
simulation \cite{gillespie1977} for the model, we show that this simple
stochastic model can predict rapid evolution phenomena in well-mixed
systems, yielding phase diagrams that are similar to those of more
complex deterministic models and in qualitative agreement with
available data. Thus key aspects of rapid evolution can be minimally modeled by
subpopulation dynamics driven simply by intrinsic demographic
stochasticity, without additional biological mechanisms. Our model can
serve as a starting point for analyzing spatial distributions and large
fluctuations such as extinction.

The physical explanation for anomalous cycles was understood early on
\cite{yoshida2003}. In contrast to the  $\pi/2$ phase shift of the
conventional predator-prey model, evolutionary cycles with a $\pi$ phase
shift can arise because of the existence of a mutant prey population
that can defend itself from the predator but which incurs a metabolic
cost. The defended prey compete with the wild type for nutrients and
thus delay the regrowth of the wild-type prey.  The resulting
additional phase lag of the wild-type prey behind the defended prey is
about $\pi/2$ because the wild-type prey must grow back before the
population of the defended prey will return to its minimum level. When
the defended prey have very effective defense without significant
metabolic cost, there is substantial delay of the regrowth of the
wild-type prey. If the wild-type prey lag the defended prey by $\pi$,
their fluctuations offset each other and thus the dynamics of the
total prey population appears in aggregate to be suppressed, leading to
the cryptic cycles.

\medskip \noindent {\it ILM for rapid evolution.\/} To model this quantitatively, consider a model
for a system composed of nutrients for the prey (N), the vulnerable
(wild-type) prey (W), the so-called \lq defended\rq\ (mutant) prey (D), and
the predator (P). The basic individual processes for them are regrowth of
nutrients, reproduction of prey, predation by predator, death and migration to the nearest site for all individuals:
\begin{eqnarray} \label{reactions}
&&\varnothing_i\xrightarrow{\textit{$b$}}N_i,\;\;
N_iR_i\xrightarrow{\textit{$\frac{c_R}{V}$}}R_iR_i,\;\;
R_iP_i\xrightarrow{\textit{$\frac{p_R}{V}$}}P_iP_i,\nonumber\\
&&S_i\xrightarrow{\textit{$d_S$}}\varnothing_i,\;\;
S_i\xrightarrow[{\langle ij\rangle}]{\textit{$\nu_S$}}S_{j}
\end{eqnarray}
where $\varnothing_i$ denotes the empty state at site $i$, $R=W,D$ is the prey index,
$S_i$ represents species $S=N,W,D,P$ at site $i$, and $V$ is an
effective coarse-grained or correlation volume in which there is no
significant population spatial variation. In ecology, $V$ is called the
patch size, and it acts as a control on the amplitude of demographic
fluctuations. Because $V$ is larger than the mean volume per
organism, we will make analytical progress by using an expansion in
inverse powers of $V$. The defended prey experiences a
smaller predation rate than the wild-type prey,
\textit{i.e.} $p_D < p_W$, and also has a smaller reproduction rate or larger degeneration rate due
to the metabolic cost for defense, \textit{i.e.} $c_W > c_D$ or $d_W < d_D$. For the
nutrients, $\nu_N$ and $d_N$ are set to zero. The corresponding
master equation that defines the time evolution of the probability
distribution of population states is
\begin{eqnarray} \label{master} &&\partial_t P(\{n_{S_i}\})=\sum_{\{n_{S_i}\}}\Big\{
b(E^{-1}_{N_i}-1)(n_{N_i}^{\text{max}}-n_{N_i}) \nonumber\\
&&+ \sum_S d_S(E_{S_i}-1)n_{S_i}+ \sum_{R}
\Big[\frac{c_R}{V}(E_{N_i}
E^{-1}_{R_i}-1)n_{N_i} n_{R_i}\nonumber\\
&&+\frac{p_R}{V}(E_{R_i} E^{-1}_{P_i}-1)n_{R_i} n_{P_i}\Big]\Big\}P(\{n_{S_i}\}),
\end{eqnarray}
where $\{\cdots\}$ denotes the set over all sites and species, the prey
index $R=W,D$, and the step operators $E_{S_i}^{\pm}$ are defined as
$E_{S_i}^{\pm}f(\{n_{S_i}\})=f(\{n_{S_i}\pm 1\})$.

\begin{figure}
\includegraphics[width=0.92\columnwidth]{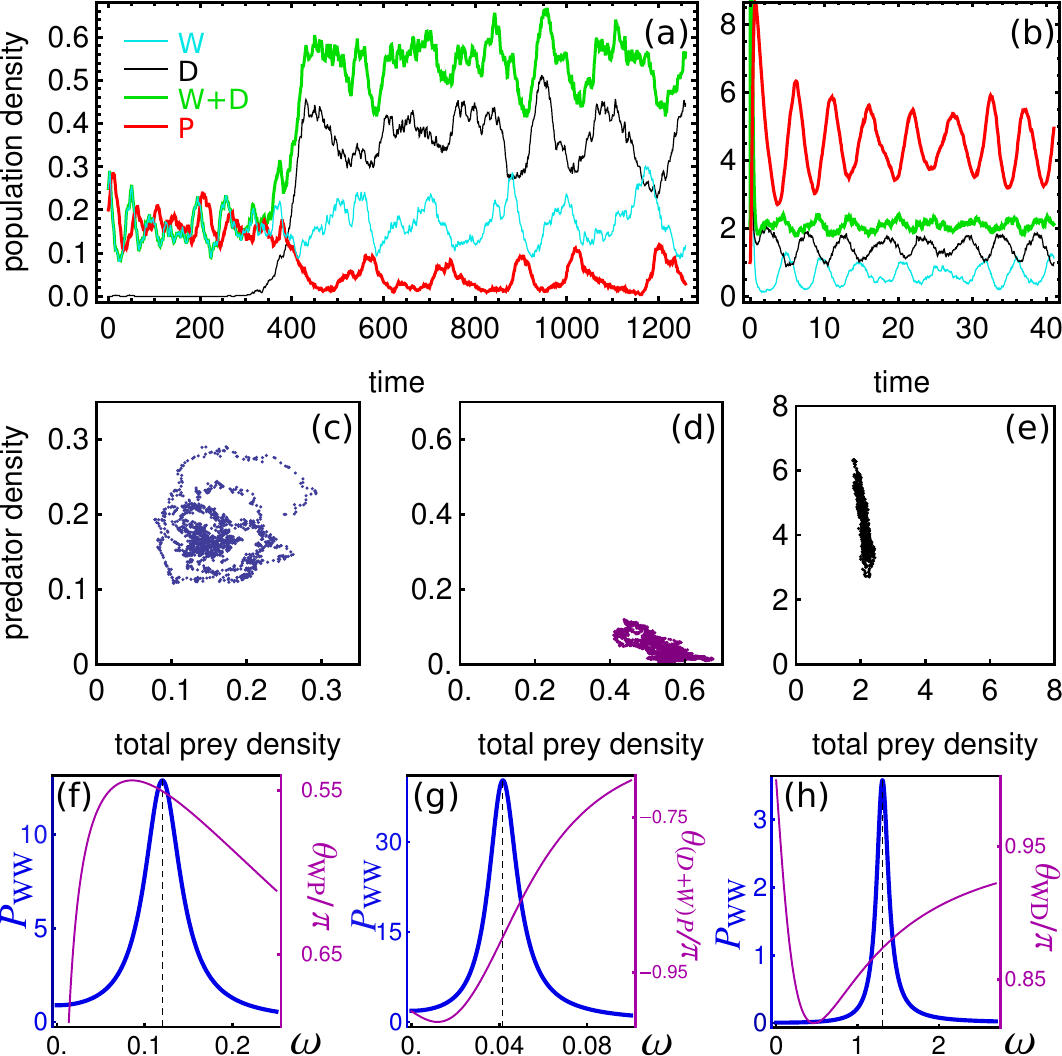}
\caption{(Color Online) Stochastic simulations for (a) evolutionary
cycles emerging from normal cycles due to random mutation and 
(b) cryptic cycles. Phase portraits of the steady states of (c) normal
cycles and (d) evolutionary cycles from the stochastic simulations show
that the phase differences between predator and the total prey
population are roughly $\pi/2$ and $\pi$, respectively, while for (e)
cryptic cycles there is no
obvious phase relationship. (f)-(h) Power spectrum of the wild-type prey
(thick curve) and phase difference spectrum (thin curve) from analytic
calculations based on ILM. The estimated phase differences are
$-0.55\pi$ and $0.905\pi$ for (f) normal cycles and (g) evolutionary
cycles, and for (h) cryptic cycles the predicted phase difference
between the wild-type prey and the defended prey is approximately
$0.874\pi$. The parameter values are (a) $V=1000$, $c_W=0.3$, $p_W=0.6$,
$c_D/c_W=0.8$, $p_D/p_W=0.01$, $d_D/d_W=1$, $\phi_{N,\text{max}}=1$,
and $b=0.1$ and (b) $V=380$, $c_W=60$, $p_W=0.92$, $c_D/c_W=0.95$,
$p_D/p_W=0.001$, $d_D/d_W=7.5$, $\phi_{N,\text{max}}=16$, and
$b=0.1$.}\label{fig1}
\end{figure}

\medskip
\noindent {\it Spatial extension.\/} To complete the specification of
the model, we need to include particle diffusion, for which the Doi
formalism \cite{doi1976} is especially convenient.  The resulting
spatially-extended model represents a non-perturbative formulation of
the model and can be used to study spatial patterns and large demographic fluctuations that
are important near the ecosystem extinction transition, where the
predator population vanishes \cite{mobilia2007,parker2010}. The procedure is to
write Eq. (\ref{master}) as a second-quantized Hamiltonian and then express the
generating functional for probabilities and correlations as a path
integral \cite{grassberger1980fock,mikhailov1981path,peliti1985path,mattis1998uses}.

Following the standard procedure, we introduce the
probability state vector in the Fock space constructed by different
occupation number states
\begin{equation}
|\psi\rangle = \sum_{\{n_{S_i}\}}P(\{n_{S_i}\})|\{n_{S_i}\}\rangle,
\end{equation}
so that the master equation becomes a Liouville equation
\begin{equation}\label{liouvillian}
\partial_t|\psi\rangle=-\hat{H}|\psi\rangle,
\end{equation}
with the Liouvillian $\hat{H}=\sum_i \hat{H}_i$
\begin{eqnarray}
\hat{H}_i&&= b\big( 1- \hat{a}_{N_i}^{\dagger} \big) \big( n_{N_i}^{\text{max}} -
\hat{a}_{N_i}^{\dagger} \hat{a}_{N_i} \big) + \sum_R \Big[ \frac{c_R}{V}
\big(\hat{a}_{N_i}^{\dagger} \hat{a}_{N_i} \hat{a}_R^{\dagger} \hat{a}_{R_i}\nonumber\\
-&& \hat{a}_{N_i} \hat{a}_{R_i}^{\dagger 2} \hat{a}_{R_i} \big) +
\frac{p_R}{V} \big( \hat{a}_{R_i}^{\dagger} \hat{a}_{R_i} \hat{a}_{P_i}^{\dagger} \hat{a}_{P_i} - \hat{a}_{R_i} \hat{a}_{P_i}^{\dagger 2} \hat{a}_{P_i} \big)\Big]\nonumber\\
+&& \sum_S \Big[ d_S \big( \hat{a}_{S_i}^{\dagger} \hat{a}_{S_i} - \hat{a}_{S_i} \big)
+ \nu_S \sum_{j\in N.N.}\big( \hat{a}_{S_i}^{\dagger} - \hat{a}_{S_{j}}^{\dagger}\big)\hat{a}_{S_i}\Big]
\label{liouvillian_whole}
\end{eqnarray}
where $\hat{a}_{S_i}^{\dagger}$ and $\hat{a}_{S_i}$ are bosonic raising
and lowering number operator for species $S$ at site $i$. Eq.
(\ref{liouvillian}) and (\ref{liouvillian_whole}) are exact and naturally allow the representation of the
many-body path integral formalism. Using the standard mapping to the
coherent-state path integral representation and applying the volume
expansion method, the
effective Lagrangian density for Gaussian-order fluctuations becomes
\begin{eqnarray} \label{L2}
\mathcal{L}^{(2)} = \tilde{\boldsymbol{\rho}}^T \partial_t \boldsymbol{\xi} -
\tilde{\boldsymbol{\rho}}^T \mathbf{A}[\{\phi_S\}]
\boldsymbol{\xi}-\frac{1}{2}\tilde{\boldsymbol{\rho}}^T \mathbf{B}[\{\phi_S\}]
\boldsymbol{\xi}
\end{eqnarray}
where $\boldsymbol{\xi}=(\xi_N,\xi_W,\xi_D,\xi_P)$ and
$\tilde{\boldsymbol{\rho}}=(\tilde{\rho}_N,\tilde{\rho}_W,\tilde{\rho}_D,\tilde{\rho}_P)$
are the fluctuation field vectors, and $\mathbf{A}$ and $\mathbf{B}$
are functions of the mean field densities $\{\phi_S\}$ given in the Supplementary Material \cite{supplementary}. Eq. (\ref{L2}) is equivalent to the Langevin
equations as a function of wavenumber $k$ and time:
\begin{eqnarray} \label{langevin}
\frac{d \boldsymbol{\xi}}{dt}&=&\mathbf{A}\boldsymbol{\xi}+ \boldsymbol{\gamma},\nonumber\\
\langle \gamma_{S}(k,t)\gamma_{S'}(k',t')
\rangle &=& \mathbf{B}_{SS'}(2\pi)^d\delta(k-k')\delta(t-t').
\end{eqnarray}
In contrast to deterministic models
\cite{fussmann2000,yoshida2003,yoshida2007,jones2007,becks2010,becks2012,bohannan1997,ellner2013,shertzer2002,yamauchi2005,cortez2010,mougi2012,yamamichi2011,mougi2012},
the dynamics depends not only on the Jacobian $\mathbf{A}[\{\phi_S\}]$
from the mean-field equation but also on the covariance matrix
$\mathbf{B}[\{\phi_S\}]$. Since $\mathbf{B}_{RR'}[\{\phi_S\}]$ in Eq.
(\ref{langevin}) is governed by the macroscopic densities, the white
noise $\boldsymbol{\gamma}$ that determines the dynamics of
fluctuations is effectively multiplicative. Without the white noise
$\boldsymbol{\gamma}$, the solutions for $\boldsymbol{\xi}$ in the
Langevin equations in Eq. (\ref{langevin}) contributed by the linear
terms are expected to decay exponentially and converge to mean-field
densities $\{\phi_S\}$. However, the multiplicative white noise plays
an important role: Whenever it can cancel out the contribution of the
eigenvalues of $\mathbf{A}$, $\boldsymbol{\xi}$ will be persistently
driven away from convergent mean-field densities, \textit{i.e.} white
noise can select the frequency in the deterministic equations,
resulting in periodic and strongly fluctuating population dynamics and
spatial patterns. This is a resonant effect induced by
demographic stochasticity through shot noise \cite{mckanePRL2005} with
the resonant frequency near the slowest decaying mode in the mean-field
solutions. Since the systems in the rotifer-algae experiments are
well-mixed, the diffusion terms are neglected in the following
calculation and simulation, but will be discussed elsewhere.

\medskip
\noindent {\it Power spectrum, phase relationship and phase diagram.}
The power spectrum of demographic noise has a resonant frequency
corresponding to the deterministic eigenvalue. The power spectrum of species
$S$, $P_{SS}(\omega)$, can be calculated by taking the Fourier transform of
the Langevin equations Eq. (\ref{langevin}),
\begin{eqnarray}
P_{SS'}(\omega)=\langle\tilde{\xi_S}(\omega)\tilde{\xi_{S'}}(-\omega)\rangle
\label{pii'}
\end{eqnarray}
and setting $S'=S$.  The Fourier transform gives the autocorrelation function,
which has the form of a polynomial of degree 6 divided by a polynomial of
degree 8, yielding a power law tail proportional to $\omega^{-2}$
at large $\omega$, as expected for quasicycles in other systems \cite{morita2005,butlerPRE2009b}.
$P_{SS}(\omega)$ peaks at a resonant
frequency that is smaller than the oscillation frequency of the
deterministic solution because of the renormalization by the white noise in
Eq. (\ref{langevin}) \cite{tauber2012}. The longer period reflects the
presence of the defended prey that causes the delay of the regrowth of the
wild-type prey and the predator.
The phase difference between the fluctuation fields is defined as
\begin{equation} \label{thetaii'}
\theta_{SS'}(\omega)=\tan^{-1}\frac{\text{Im}[P_{SS'}(\omega)]}{\text{Re}[P_{SS'}(\omega)]}\;.
\end{equation}
The phase difference between total prey and the predator,
$\theta_{(W+D)P}$, can be calculated from
$P_{(W+D)P}(\omega)=\langle(\tilde{\xi_W}(\omega)+\tilde{\xi_D}(\omega))\tilde{\xi_P}(-\omega)\rangle=P_{WP}(\omega)+P_{DP}(\omega)$.

The results of analytic calculations and simulations based on Eq.
(\ref{reactions}) are shown in Fig. \ref{fig1}. We use the Gillespie
algorithm \cite{gillespie1977} for stochastic simulations and introduce
random mutation from the wild-type prey to the defended prey. The
mutation is added purely to seed a new sub-population to see the
dramatic impact of the fixed sub-population after mutations, but plays
no significant role in the subsequent dynamics; thus mutations are
neglected in our analytical calculations below. The subsequent
anomalous dynamics due to the presence of this sub-population is
conventionally called evolution in the ecological literature, because
the presence of the additional strain emerges from mutation, and we are
interested in following the frequency in the population of the mutant
strain. We tried to simulate the experimental results of the
rotifer-algae chemostat, where the control parameters are the nutrient
concentration in flow media, $\phi_N^{\text{max}}$, and the dilution
rate, $b$. The natural degradation rates of the wild-type prey and
predator are assumed to be much slower than the dilution rate, and
therefore $b\approx d_P\approx d_W < d_D$ (the defended prey is less
healthy). In Fig. \ref{fig1}(a), at first there are only the wild-type
prey and the predator in the system, and the dynamics exhibits normal
cycles where the predator lags behind the prey by $\pi/2$. When
predation pressure is high, around $t \sim 400$, a mutation has given
rise to a defended prey population that subsequently adapts to
dominate the population and cause additional delay in growth of the
wild-type prey and the predator, leading to evolutionary cycles with a 
$\pi$ phase shift between the total prey and the predator. Fig.
\ref{fig1}(b) shows an example of cryptic cycles, where the defended
prey has a similar reproduction rate as that of the wild-type prey,
\textit{i.e.} $c_D \sim c_W$, and the defended prey can advance the
wild-type prey by nearly $\pi$ and thus the total prey population is
suppressed. The quasicycle calculations in Fig. \ref{fig1}(f)-(h) for
the power spectrum and the phase spectrum
well predict the simulation results in Fig.
\ref{fig1}(c)-(e). Besides the expected randomness in the dynamics from
the stochastic simulation, Fig. 1(a) and (b) also show similar
asymmetric profiles and the longer period after the subpopulation
emerges, as in the experimental data in
\cite{fussmann2000,yoshida2003,yoshida2007,jones2007,becks2010,becks2012,ellner2013}.

\begin{figure}
\includegraphics[width=1.0\columnwidth]{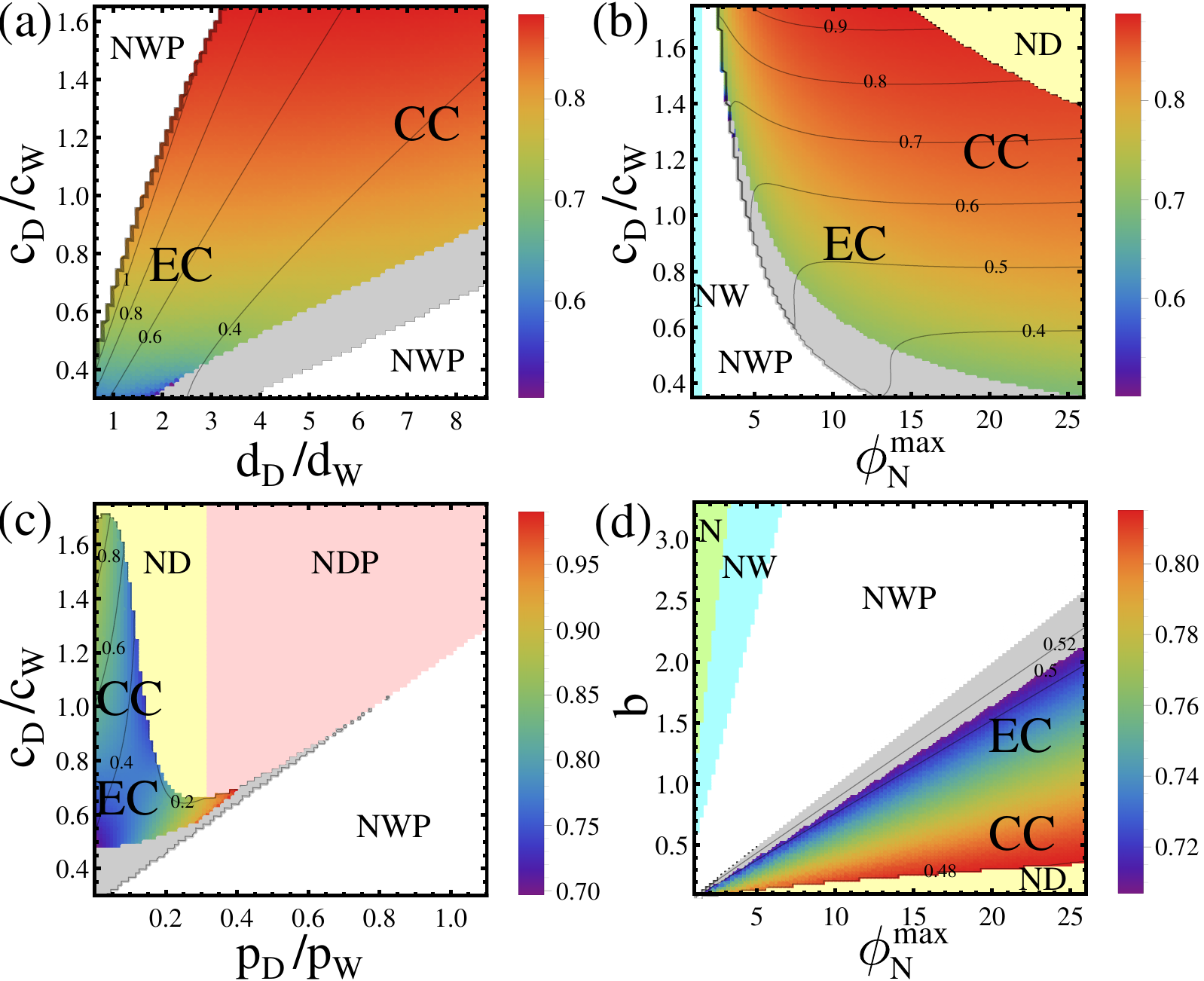}
\caption{(Color Online) Phase diagrams for evolutionary cycles (EC) and cryptic cycles
(CC) calculated from ILM with respect to the ratio of the prey reproduction
rate ($c_D/c_W$), the ratio of the predation rate ($p_D/p_W$), the maximum
nutrient concentration ($\phi_N^{\text{max}}$) and the dilution rate
($b$). The gradient-colorful region corresponds to the coexistence of
all species, and in the other regions the rapid
evolution is not stable, with corresponding letters indicating the
coexistence of only certain species. The coexistence states are decided
by the mean-field densities and their ratio to the fluctuations; when
fluctuations are larger than mean-field solutions, the dynamics is
under high risk of extinction. The color legend represents the
predicted phase difference between the wild-type prey and the defended
prey ($\theta_{WD}$) for rapid evolution, in units of $\pi$. The
contours are the estimated amplitude ratios of wild-type prey to
predator, indicating the tendency to be cryptic cycles. In the gray
region near transition, the two types of prey start to decouple,
leading to degenerate peaks in power spectra, and thus the phase is not
well-defined. Except for the axis specified in each diagram, the
parameters in the calculations are $V=300$, $c_W=1$, $p_W=1$,
$c_D/c_W=0.8$, $p_D/p_W=0.01$, $d_D/d_W=3.5$,
$\phi_{N}^{\text{max}}=16$, and $b=0.6$. The predicted phase diagram is
consistent with stochastic simulation.} \label{fig2}
\end{figure}

The phase diagram is usually studied by linear stability analysis of the mean
field equations (for example, see Eq. (7)-(9) in the Supplementary Material \cite{supplementary}). To reduce the dimension of
parameter space, variables are rescaled to be dimensionless: $\bar{t}\equiv
bt,\bar{d}_S\equiv d_S/b,\bar{\phi}_S\equiv
/\phi_N^{\text{max}},\bar{c}_S\equiv c_S \phi_N^{\text{max}} /b$ and
$\bar{p}_S\equiv p_S\phi_N^{\text{max}}/b$.
However, this rescaling is rather subtle in stochastic
calculations. For example, matrices $\mathbf{A}$ and
$\mathbf{B}$ from Eq. (\ref{langevin}) scale with
$1/\phi_N^{\text{max}}$ as mean-fields $\phi_S$, but
$\mathbf{\gamma}$ in Eq. (\ref{langevin}) rescales with
$1/\sqrt{\phi_N^{\text{max}}}$, resulting in
\begin{equation}
\frac{\xi_S}{\phi_S} \sim \frac{1}{\sqrt{\phi_N^{\text{max}}}}\frac{\bar{\xi}_S}{\bar{\phi}_S}
\end{equation}
where $\bar{\xi}_S$ are the rescaled demographic noise fields.
Therefore, for two stochastic individual-level models with the same
mean-field limit after rescaling, demographic fluctuations are more
important in the model with smaller nutrient carrying capacity
$V\phi_N^{\text{max}}$. Thus neglecting fluctuations as in the
conventional rescaling for mean-field equations can potentially cause
unphysical predictions for the phase diagram. To avoid this situation,
we examine the stability of solutions by comparing the amplitude of the
lowest order population fluctuations with their mean fields.

Fig. \ref{fig2} shows the calculated phase diagrams of ILM in Eq.
(\ref{reactions}). In Fig. \ref{fig2}(a), due to the cost for defense,
the defended prey have inferior reproduction rate ($c_D < c_W$) or are
unhealthier than the wild-type prey ($d_D > d_W$), leading to
evolutionary cycles (EC). When the cost of reproduction is low, cryptic
cycles (CC) can occur, where $\theta_{WD}\approx \pi$. If $c_D$ is
moderate, it is possible to have a correspondingly high death rate, and
thus the fluctuations of prey are suppressed relative to the wild-type
prey, causing the dynamics to be cryptic. In Fig. \ref{fig2}(b), under
high $\phi_N^{\text{max}}$, the defended prey are more likely to grow
and dominate the system, which causes the wild-type prey to experience
a greater phase lag than the defended prey, and the dynamics tends
towards a completely cryptic cycle. In Fig. \ref{fig2}(c), if $p_D$ is
low, then higher $c_D$ can lead to more phase delay and thus gives
cryptic cycles. When $p_D$ increases, the predator has greater food
resource available from the defended prey, yielding a larger
population, which then consumes more of the wild-type prey; this in
turn reduces the wild-type prey population and leads to the dominance
of the defended prey. In such a situation, the wild-type prey
experiences a greater phase delay (nearly $\pi$) behind the defended
prey, but the wild-type prey population is too small to cancel out the
fluctuations of the defended prey population, and thus the dynamics
cannot be characterized as cryptic. Our result in Fig. \ref{fig2}(c)
predicts a similar but slightly different phase diagram to Fig. 3 in
\cite{yoshida2007}; the region where all species coexist as
predicted by the stochastic model is smaller than the deterministic
solutions, because of extinction fluctuations near phase boundaries. In Fig.
\ref{fig2}(d), under small $b$, \textit{i.e.} slow supplement of the
nutrient and low reduction rate from dilution, although both
subpopulations of the prey have low reproduction, the wild-type prey
population decreases more due to predation while the defended prey has
a greater chance to compete for nutrient; thus the system is more
likely to show cryptic cycles.

Our results show that rapid evolution strongly renormalizes the
ecosystem time scale, and the prediction of the coexistence region can
help estimate the risk of extinction and the impact of the rate of
environmental changes (for example, the dilution rate and nutrient
concentration in the rotifer-algae system). Our model can also be used
to study spatial-extended situations in natural ecosystems or
laboratory experiments that are not in a well-mixed chemostat.

In summary, we have shown clearly that a generic stochastic
individual-level model can yield rapid evolution phenomena, and that
anomalous dynamics can arise without special assumptions or fine
tuning, in sharp contrast to existing results in the ecology literature
based on deterministic models. We expect this description to be
especially useful to study the transition to rapid evolution from
normal cycles since before the transition the mutant prey population
has low relative abundance and is thus likely to exhibit strong effects
of demographic stochasticity and spatiotemporal fluctuations.

\emph{Acknowledgements.} We thank S.P. Ellner and U. T\"{a}uber for helpful discussions. We thank P. Rikvold for bringing Ref \cite{murasePRE2010} to our attention after the completion of this work. This material is partially supported by the National Aeronautics and Space Administration through the NASA Astrobiology Institute under Cooperative Agreement No. NNA13AA91A issued through the Science Mission Directorate.

\bibliographystyle{apsrev4-1}
\bibliography{Rapid_evolution_ref}


\appendix

\begin{center}
{\bf SUPPLEMENTARY MATERIAL}
\end{center}

\section{Path integral formalism for rapid evolution}
\label{sec:ssec1}

By using the coherent-state representation, the Hamiltonian can be mapped
onto the basis of coherent states and becomes a function of $\alpha_S^*$ and
$\alpha_S$ which are the left and right eigenstates of $\hat{a}_S^{\dagger}$ and $\hat{a}_S$
respectively for species $S=N,W,D,P$. Since in general multiple individuals can occupy the same site in spatial extended systems, $\hat{a}_S^{\dagger}$ and $\hat{a}_S$ are considered to follow the bosonic commutation relation. The effective Lagrangian density in the path integral
becomes
\begin{eqnarray} \label{lagrangian}
\mathcal{L}= \sum_i \left[ \alpha_S^* \left( \partial_t - \nu_S \nabla^2 \right) \alpha_S + H \left(\right\{
\alpha_S^* \left\}, \right\{ \alpha_S \left\} \right)\right]
\end{eqnarray}
where $\nu_N\equiv 0$.

To study the fluctuations about the mean-field densities, it is convenient to map the system from field variables onto the physical variables by applying the canonical Cole-Hopf transformation \cite{mobilia2007}
\begin{equation}
\alpha_S^*=e^{\tilde{\rho}_S}\;\;, \alpha_S=\rho_S e^{-\tilde{\rho}_S}
\end{equation}
where $\rho_S$ are the population variables for species $S$ and $\tilde{\rho}_i$ are analogous to fluctuation variables. The Hamiltonian density under the
transformation is obtained as
\begin{eqnarray}
H&=&b\left(1-d^{\tilde{\rho}_N}\right)\left(n_{N,\text{max}} - \rho_N \right)
+ \frac{c_R}{V}\left(1 - e^{\tilde{\rho}_R - \tilde{\rho}_N} \right)
\nonumber\\
&+& \frac{p_R}{V}\left(1 - e^{\tilde{\rho}_P - \tilde{\rho}_R} \right) + d_R
\rho_R \left( 1 - e^{-\tilde{\rho}_R} \right) \nonumber\\
&+& d_P \rho_P \left( 1 - e^{-\tilde{\rho}_P} \right).
\end{eqnarray}
Further we apply the ansatz
\cite{butlerPRE2009}
\begin{equation} \label{ansatz}
\tilde{\rho}_S \rightarrow \frac{\tilde{\rho}_S}{\sqrt{V}},\;\;\; \rho_S =
V\phi_S + \sqrt{V} \xi_S,
\end{equation}
where $\langle\tilde{\rho}_S\rangle$ are the mean-field population
density variables and the deviations around them, $\phi_S$, are of
order $1/\sqrt{V}$. The patch size $V$ becomes the system size in the
well-mixed case. This expansion will lead to a lowest order of
fluctuations in Gaussian form. After applying the expansion in Eq.
(\ref{ansatz}), the Lagrangian density in Eq. (\ref{lagrangian}) can be
separated into different orders of $\sqrt{V}$
\begin{equation}
\mathcal{L}=\sqrt{V}\mathcal{L}^{(1)} + \mathcal{L}^{(2)} +...
\end{equation}
Here
\begin{eqnarray}\label{L1}
\mathcal{L}^{(1)} &=& \sum_S \tilde{\rho}_S \partial_t \phi_S + b\phi_N
\tilde{\rho}_N + \sum_R \big[ - \nu_R \tilde{\rho}_R \nabla^2 \phi_R \nonumber\\
&+& c_R \phi_N \phi_R \left( \tilde{\rho}_N - \tilde{\rho}_R
\right)
+ p_R \phi_R \phi_P \left( \tilde{\rho}_R - \tilde{\rho}_P \right)\nonumber\\
&+& d_R
\phi_R \tilde{\rho}_R \big]
+ d_P \phi_P \tilde{\rho}_P - \nu_P \tilde{\rho}_P \nabla^2 \phi_P.
\end{eqnarray}
The stationary solution from $\frac{\delta\mathcal{L}_1}{\delta\tilde{\rho}_S}=0$ gives the mean-field dynamics:
\begin{eqnarray} \label{mean-field-eq}
&& \partial_t\phi_N = b\left(\phi_{N,\text{max}}-\phi_N\right)-c_R\phi_N\phi_R,\;\;\;\;\;\;\\
&& \partial_t\phi_R = \nu_R \nabla^2 \phi_R +
c_R\phi_N\phi_R - p_R\phi_R\phi_P - d_R\phi_R,\;\;\;\;\;\;\\
&& \partial_t\phi_P = \nu_P \nabla^2 \phi_P + \sum_R
p_R\phi_R\phi_P - d_P\phi_P.\;\;\;\;\;\;
\end{eqnarray}

The Lagrangian density in the next order is 
\begin{eqnarray} 
\mathcal{L}^{(2)} = \tilde{\boldsymbol{\rho}}^T \partial_t \boldsymbol{\xi} -
\tilde{\boldsymbol{\rho}}^T \mathbf{A}[\{\phi_S\}]
\boldsymbol{\xi}-\frac{1}{2}\tilde{\boldsymbol{\rho}}^T \mathbf{B}[\{\phi_S\}]
\boldsymbol{\xi},\;\;\;\;\;\;\;\;
\end{eqnarray}
where $\boldsymbol{\xi}=(\xi_N,\xi_W,\xi_D,\xi_P)$ and $\tilde{\boldsymbol{\rho}}=(\tilde{\rho}_N,\tilde{\rho}_W,
\tilde{\rho}_D,\tilde{\rho}_P)$ are the fluctuation field vectors, and
\begin{eqnarray}
&&\mathbf{A}_{NN}=-b-c_R\phi_R,\;\mathbf{A}_{NR}=-\mathbf{A}_{RN}=-c_R\phi_N,\nonumber\\
&&\mathbf{A}_{NP}=\mathbf{A}_{PN}=\mathbf{A}_{WD}=\mathbf{A}_{\nu_R}=0,\nonumber\\
&&\mathbf{A}_{RR}=-\nu_R k^2 + c_R\phi_R-p_R\phi_P-d_R,\nonumber\\
&&\mathbf{A}_{RP}=-\mathbf{A}_{PR}=-p_R\phi_R,\; \mathbf{A}_{PP}= - \nu_P k^2 + p_R\phi_R-d_P, \nonumber\\
&&\mathbf{B}_{NN}=b(\phi_{N}^{\text{max}}-\phi_N)+c_R\phi_N\phi_R,\nonumber\\
&&\mathbf{B}_{NR}=\mathbf{B}_{RN}=-c_R\phi_N\phi_j,\nonumber\\
&&\mathbf{B}_{NP}=\mathbf{B}_{PN}=\mathbf{B}_{WD}=\mathbf{B}_{DW}=0,\nonumber\\
&&\mathbf{B}_{RR}= \nu_R \phi_R k^2 + c_R\phi_N\phi_R + p_R\phi_R\phi_P + d_R \phi_R,\nonumber\\
&&\mathbf{B}_{RP}=\mathbf{B}_{PR}= - p_R\phi_R\phi_P,\nonumber\\
&&\mathbf{B}_{PP}= \nu_P \phi_P k^2 + p_R\phi_R\phi_P + d_P \phi_P.
\end{eqnarray}

Following the Martin-Siggia-Rose response function formalism
\cite{martin1973statistical,janssen1976lagrangean}, the next order
$\mathcal{L}^{(2)}$ generates a Langevin equation capturing the
dynamics with Gaussian fluctuations. However, it is important to
emphasize that the Gaussian noise here is intrinsic because these
fluctuations originate from the demographic stochasticity of the
population at each time step.  The outcome is quasicycles induced by a
resonant amplification of intrinsic fluctuations \cite{mckanePRL2005},
which leads to a slowly-decaying tail in the power spectrum that is
distinct from the behavior of limit cycles with additive noise
\cite{butlerPRE2011}. In our calculation, the noise depends on the mean
field values because of the linearization step performed as part of the
van Kampen volume expansion \cite{kampen1961power}. However, we
emphasize that this is the signature of multiplicative noise within the
van Kampen framework.  If we had chosen to perform a Kramers-Moyal
expansion \cite{kramers1940brownian,moyal1949stochastic} instead, the
noise would have been manifestly multiplicative, and linearization to
obtain a systematic calculational procedure would have arrived at the
van Kampen expansion results presented here.

\begin{figure}
\includegraphics[width=1\columnwidth]{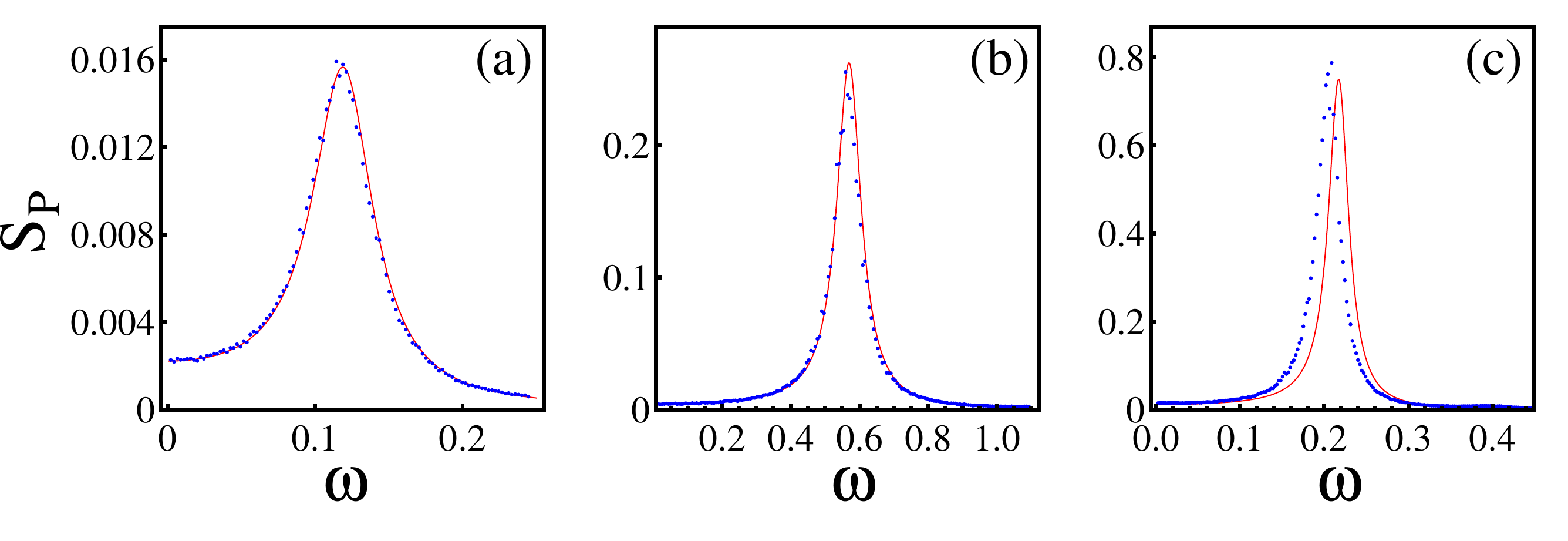}
\caption{Examples of comparison between analytic calculation (red curve) and stochastic simulation (blue dots) of power spectrum of population fluctuations of predator for (a) normal cycles, (b) evolutionary cycles and (c) cryptic cycles
in individual level model. The parameters in
calculations and simulations are (a) $V=2000$, $b=0.1$, $c_W=0.3$, $p_W=0.6$, $\phi_{N}^{\text{max}}=1$, (b) $V=1600$, $b=0.6$, $c_W=1$, $p_W=1$, $\phi_{N}^{\text{max}}=5$, $c_D/c_W=1.6$,
$p_D/p_W=0.001$, $r_D/r_W=3.5$ and (c) $V=1600$, $b=0.1$, $c_W=60$, $p_W=0.92$, $\phi_{N}^{\text{max}}=16$, $c_D/c_W=0.95$,
$p_D/p_W=0.001$, $r_D/r_W=7.5$.}
\label{fig1_SM}
\end{figure}

\begin{figure}
\includegraphics[width=0.55\columnwidth]{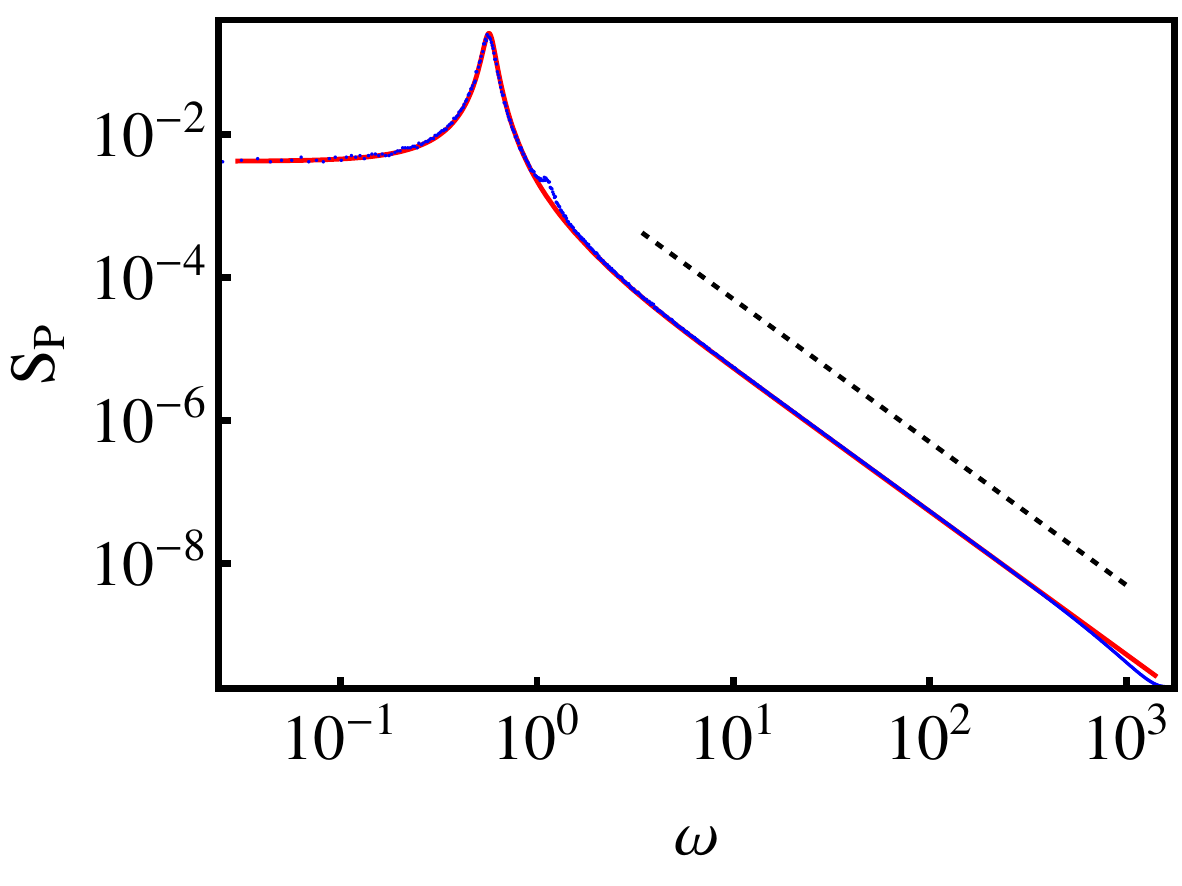}
\caption{The logarithm scale of comparison of power spectrum between analytic calculation and stochastic simulation in Fig \ref{fig1_SM}(b). The tail with $\omega^{-2}$ scaling, indicated by a reference dotted line with slope of $-2$, is the signature of quasicycles and is predicted by the analytic calculation based on individual level model.}
\label{fig2_SM}
\end{figure}

\section{Comparison between analytic calculation and stochastic simulation}
\label{sec:ssec1}

We have computed the power spectra, and compared the results with the stochastic simulation. The power spectrum for species $S$ is calculated analytically by $S_{S}(\omega)=P_{SS}(\omega)=\langle\tilde{\xi}_S(\omega)\tilde{\xi}_S(-\omega)\rangle$, and from Gillespie stochastic simulations of the ILM, using the formula
\begin{eqnarray}
S_{S}(\omega)=\frac{1}{T}\langle\tilde{{\xi}'}_S(\omega_m)\tilde{{\xi}'}_S(-\omega_m)\rangle,
\end{eqnarray}
where $T$ is the duration of total $N$ samplings with discrete time $t_n=(n-1)\Delta t$ and the discrete Fourier transform of $\xi$ is defined as
\begin{eqnarray}
\tilde{{\xi}'}(\omega_m)&=&\sum_{n=1}^{N} \xi(t_n)e^{i\omega_m t_n} \Delta t \nonumber\\
&=&\frac{T}{N}\sum_{n=1}^{N} \xi(t_n)e^{i 2\pi (m-1)(n-1)/N}.
\end{eqnarray}

The peaks and magnitudes of the power spectra of calculation and simulation have good agreement when the Gaussian approximation is valid. There are slight deviations when either the wild-type prey or predator has a small population size. In such a case, the dynamics of fluctuations is dominated by the species with smaller population, leading to a skewed and leptokurtic distribution of population fluctuations. Such suppressed fluctuation distribution can explain the deviation of the power spectra of simulation data from the analytic calculation when there is large discrepancy in population sizes between species. Fig. \ref{fig1_SM} shows examples of comparison between analytic calculation and stochastic simulation for normal cycles, evolutionary cycles and cryptic cycles. When population sizes are similar for each species and are not small, analytic calculation based on Gaussian fluctuations provides good quantitative prediction of characteristic frequency and the shape of power spectrum. Fig. \ref{fig2_SM} shows the $\omega^{-2}$ scaling in power spectrum as the signature of quasicycles is also captured by analytic calculation. For cryptic cycles where the oscillations of predator population are relatively larger and the prey population size is usually smaller, the Gaussian approximation is expected to have less quantitative agreement and underestimate the amplitude of fluctuations as shown in Fig. \ref{fig1_SM}(c).

\end{document}